# Project portfolio planning in the pharmaceutical industry – strategic objectives and quantitative optimization

Stig Johan Wiklund[1], Magnus Ytterstad[1], Frank Miller[2]

[1]Captario AB, Göteborg, Sweden
[2]Linköping University, Linköping, Sweden

E-mail: stig-johan.wiklund@captario.com

## Abstract

Many pharmaceutical companies face concerns with the maintenance of desired revenue levels. Sales forecasts for the current portfolio of products and projects may indicate a decline in revenue as the marketed products approach patent expiry. To counteract the potential downturn in revenue, and to establish revenue growth, an in-flow of new projects into the development phases is required. In this article, we devise an approach with which the in-flow of new projects could be optimized, while adhering to the objectives and constraints set on revenue targets, budget limitations and strategic considerations on the composition of the company's portfolio.

## Introduction

It could be claimed that an ultimate goal for a privately held company is always to generate maximum value to its owners (e.g. Brătianu & Bălănescu, 2008). In terms of optimizing the selection of projects for a portfolio, this goal might be translated into maximizing the net present value (NPV), or other similar value metric, of the portfolio. The real situation is however often more complex and would typically involve other considerations. The pharmaceutical companies' contributions to the health of the population (Leisinger 2005), reduction of therapy costs to society (LaMattina, 2022), and their capacity to provide treatments to patients in need (Baird et al., 2014), are important aspects of the activities of the pharmaceutical industry. The robustness of the portfolio, as represented by a diversity of projects across the different phases of development, and across various disease areas, are also important aspects of a healthy drug development portfolio.

Many authors have contributed to the topic of optimizing the design of drug development programs and how to optimally select a subset of available projects while adhering to the available budget. Examples include He et al (2021) focusing on proof-of-concept studies, Patel et al (2013) emphasizing the design of phase 3 programs, Farid et al (2021) developed optimization methods for portfolio selection, whereas Graham et al (2020) focused on the methodological aspects and conducted a comparison of optimization methods. Dimitri (2011) asks, in the title of his paper, a question that most portfolio managers and strategic decision makers would probably have been confronted with: "What constitutes an optimal portfolio of pharmaceutical compounds?" In an attempt to find an answer to this question, three different economic goals were postulated: (i) maximizing the probability of registering at least one compound in the portfolio, (ii) maximizing the expected profit from the portfolio, and (iii) maximizing expected productivity from the portfolio.

Nadar et al. (2015) suggest a method that adjusts the reward and risk profiles of each project in the product development pipeline as they move through the development process. Perez-Escobedo et al. (2012) takes a multi-objective approach using genetic algorithms to solve a stochastic optimization model. Hassanzadeh et al. (2014) discuss the project selection problem from the perspective of a contract research organization. David and Belogolovsky (2019) argue that pharmaceutical companies often rely on cycles of R&D portfolio reviews to decide how resources should be allocated among the currently available projects. In response to this they propose an approach whereby more strategic thinking is brought into the R&D portfolio planning. A comprehensive overview of different perspectives on the topic was provided by Antonijevic (2015), editing the book entitled Optimization of Pharmaceutical R&D Programs and Portfolios.

While we do appreciate the relevance of all the mentioned research, we want to emphasize right from the outset that the subject of this article has a different focus than most of the existing literature. We do not address the issue of how to improve ongoing projects or optimize the currently available portfolio, but we do instead focus on the addition of new projects that is required for the future to maintain a portfolio with long-term desired properties. Our contribution does not lie in the short-term enhancement of current projects, but it offers an approach for strategic decision-makers to get a quantitative view of the long-term prospects of the portfolio and the in-flow required to achieve company goals. Instead of selecting from currently available projects, we will be following the ideas outlined by Wiklund et al. (2023) and focus our approach on how to best add the projects needed to replenish the portfolio for the future.

Currently marketed projects may encounter increased competition, or they may lose their patent exclusivity, and projects in development may fail and be terminated before reaching the market and generating revenue. Additionally, the company may have an objective to achieve a growth in revenue, above the current level. For all these reasons, it may be required that new projects are to be added to the development portfolio. Setting the R&D organization up for producing new drug candidates for development is a complex undertaking, requiring large resources, many years of focused work and considerate planning. The current article intends to provide a framework for quantitatively supporting this planning process.

Wiklund et al (2023) highlighted the fact that the strategic objective for optimizing a development portfolio could be defined in many different ways. The objective could focus on achieving a certain revenue target over time, while keeping the development cost as low as possible. Framed differently, the objective could be to ensure that development costs are contained within an available budget, and to maximize future revenue given the budget. A summary of some different objectives is given in Table 1. The notation 1A-4B, representing the different objectives, follows the notation presented in Wiklund et al (2023). The optimization problem corresponding to the different objectives of Table 1 will be defined in more detail in a later section.

Table 1. Overview of some alternatives of framing the objective for the optimization of drug development portfolios.

| Strategic focus | Variable to optimize while achieving strategic focus | | |
|---|---|---|---|
| | | Minimize cost | |
| | | Minimize mean cost over planning period | Minimize the highest cost relative to budget during planning period |
| **Ensure that Revenue>X** | Ensure that mean of revenue is >$X_1$ over planning period | **1A** | **1B** |
| | Ensure that lowest revenue during planning period is >$X_2$ | **2A** | **2B** |
| | | Maximize revenue | |
| | | Maximize mean revenue over planning period | Maximize lowest revenue relative to target during planning period |
| **Ensure that Cost<Y** | Ensure that mean of cost is <$Y_1$ over planning period | **3A** | **3B** |
| | Ensure that highest cost during planning period is <$Y_2$ | **4A** | **4B** |

In addition to the objectives and constraints that are resulting from considerations on revenue targets and cost budgets, there are obviously other considerations of importance to a drug development portfolio. To achieve a balanced portfolio Elbok and Berrado (2017) argue that balance constraints are considered in addition to the overall goal of value maximization. Suggested dimensions for diversifications are long vs short term gain, low vs high risk, or breakdown by market segments. The low vs high risk balance can be represented by the novelty of the drugs being developed, where low risk projects use proven API's and high risk projects use novel or unproven compounds (Ding et al., 2014). An approach commonly used in industry is to divide the resource budget into smaller focused budgets that are dedicated for specific strategies. Chao and Cavadias (2008) argue that this approach ensures access to resources for projects that would otherwise not be selected, if selection was based entirely on common project valuation methods such as net present value (NPV). We will in this paper develop an approach to portfolio planning and optimization, in which the portfolio composition may be quantitatively guided by these types of criteria. We will consider the case where the company during a planning period is striving to achieve a balanced portfolio of projects, by

- having at least $F_i$ projects in development Phase $i$
- having at least $G_j$ projects in development in disease area $j$
- launching to the market at least $H_j$ projects from disease area $j$

It should be noted that the modelling and optimization approach outlined in this paper is generic and would allow for a wide range of constraints of this type. We have focused the presentation on these examples of constraints as we think that they are representative of considerations that are often relevant for mid- to large-sized pharmaceutical companies. Having a balanced number of projects in each phase of development is important in order to obtain continuity in R&D operations and delivery. A presence across disease areas is important to reduce scientific and competitive risks. Should a disruptive breakthrough occur in a certain disease area, it is important that the company could mitigate impact by presence in other areas. A steady flow of launched products is of course also important. It is necessary to replenish the marketed portfolio, replacing older drugs that lose their market exclusivity, but regular successful market launches may also be important to maintain investor confidence in the company. The objective of this paper is to devise a methodology, by which the future addition of projects to the company's portfolio could be optimized, while adhering to constraints that the company might specify, as described above.

The rest of the article is structured as follows. In the next section, we present a quantitative model for development projects, representing the future additions to the company's portfolio. We then present the methods used to perform the actual optimization. An illustrating example is given to clarify the intended use of the proposed methodology. A discussion section concludes the paper.

# A template model for drug development projects

## Model and assumptions

The methodology is presented by applying a generic model for a drug development project, as schematically illustrated in Figure 1. Key attributes of this model are the duration of each phase, $D_i$, and the cost incurred at each phase, $C_i$, for $i \in \{1,2,3,r\}$. The probability that the project successfully proceeds from Phase $i$ to the next phase is denoted $P_i$, and projects successfully reaching the market will generate revenue, $R$. These key attributes will be further detailed below.

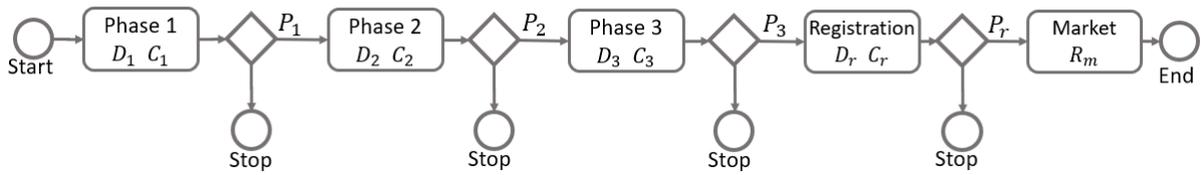

*Figure 1. Schematic illustration of a generic drug development process.*

To evaluate the expected future composition of the portfolio, we define the probability that the project is active in Phase $i$ at a given time $t$. The probability that a project is active in Phase 1 and Phase 2, respectively, are given by

$$\pi_{1t} = \begin{cases} 1 & 0 \leq t \leq D_1 \\ 0 & otherwise \end{cases} \qquad \pi_{2t} = \begin{cases} P_1 & D_1 < t \leq D_1 + D_2 \\ 0 & otherwise \end{cases}$$

and similarly for $\pi_{3t}$ and $\pi_{rt}$. In more formal notation, the probability of each phase is

$$\pi_{it} = \prod_{0 \leq i' \leq i-1} P_{i'} \qquad for\ t\ in\ the\ interval \quad \sum_{0 \leq i' \leq i-1} D_{i'} < t \leq \sum_{1 \leq i' \leq i} D_{i'}$$

where $P_0 = 1$ and $D_0 = 0$. We assume here, for ease of notation, that the project starts at time zero.

The overall probability that a project is still active in development at time $t$ is the sum of the functions for each phase, i.e. $\pi_t = \sum_i \pi_{it}$. If the phase durations, $D_i$, are considered known and fixed, the probability $\pi_t$ would be a decreasing step function. More realistically, the future phase durations are unknown, and the corresponding uncertainty could be given in the model as a stochastic distribution. In this case, $\pi_t$ would be represented by a continuous function over time. A schematic illustration is given in Figure 2.

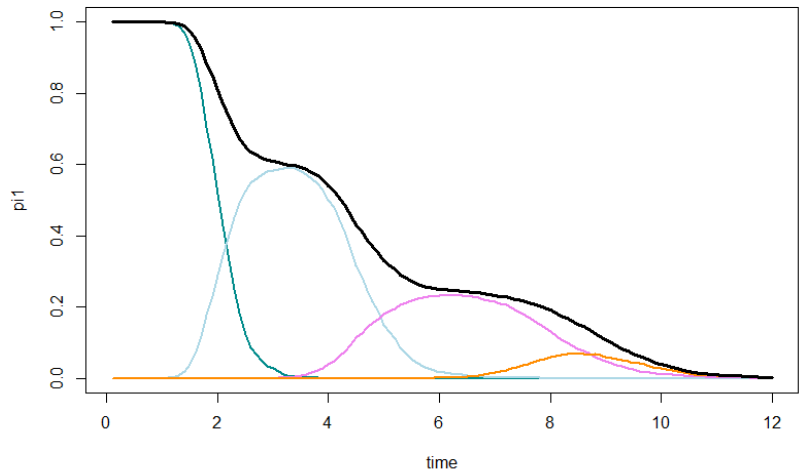

*Figure 2. An illustrating example of the probability, $\pi_t$, (bold black line) that a project is active in development over time. The thinner colored lines show the probabilities $\pi_{1t}$, $\pi_{2t}$, $\pi_{3t}$, $\pi_{rt}$, for being active in Phase 1, 2, 3, or registration, respectively (from left to right).*

The expected (risk-adjusted) cost incurred by a project at time $t$ is given by $\Gamma_t = \sum_i C_i \pi_{it}$, i.e. the cost of each phase weighted by the probability of being in the phase at time $t$. If the phase costs and durations were known, the expected cost, $\Gamma_t$, would be a step function. However, we will assume that the costs, as well as the durations, are unknown and represented by stochastic distributions. In this case the expected cost may be a continuous function as schematically illustrated in Figure 3.

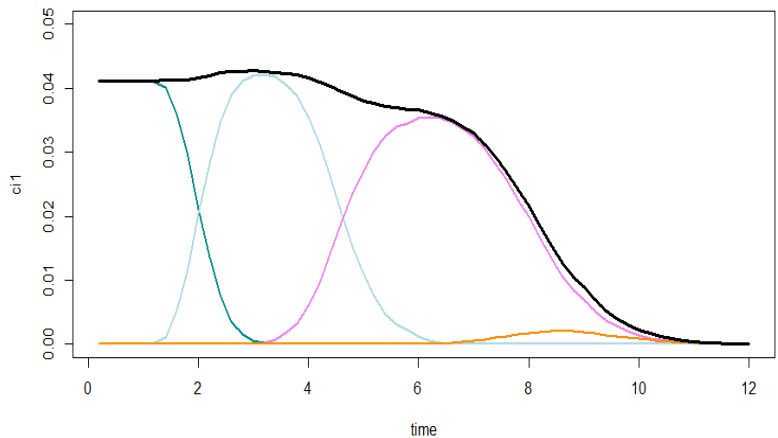

*Figure 3. An illustrating example of the risk-adjusted cost, $\Gamma_t$, (bold black line) incurred by a project over time. The thinner colored lines show the expected (risk-adjusted) costs $C_1\pi_{1t}$, $C_2\pi_{2t}$, $C_3\pi_{3t}$, $C_r\pi_{rt}$, for Phase 1, 2, 3, or registration, respectively (from left to right).*

The presentation has so far implicitly assumed that all projects are expected to be represented by the same model. It is however obvious that a typical portfolio comprises projects of different categories and properties. We will in this article focus on projects belonging to different disease areas, but the concept is generic and could as well be adapted to other categorizations of projects.

Let *j* denote the disease area for which a project is developed. The project model input would then be given separately for each disease area as $D_{ij}$, $C_{ij}$ and $P_{ij}$. Correspondingly, the definition of the risk-adjusted probabilities ($\pi_t$ and $\pi_{it}$) could be made separately for each disease area, yielding the disease-specific parameters $\pi_{jt}$ and $\pi_{ijt}$.

## Current development portfolio

An obvious starting point for the analysis of the composition of a future portfolio is the current state of the company's development portfolio. Let $K_{ij0}$ denote the number of projects that the company has, at the time of analysis, in Phase *i* in disease area *j*.

The probability that such a proj0ect will remain in development at time *t* will be denoted $\pi_{jt}^i$. For a project currently in Phase 2, we will have $\pi_{jt}^2 = P_1^{-1} \pi_{jt^*}$, where $t^* = t - D_1 - D_2/2$. This corresponds to taking the original probability function, $\pi_{jt}$, and calibrate it to equal one in the middle of Phase 2, since we know that this project has successfully transitioned from Phase 1. An illustration of this case is given in Figure 4. The corresponding general definition for projects currently in any phase *i* of development is

$$\pi_{jt}^i = \left( \prod_{i'<i} P_{i'} \right)^{-1} \cdot \pi_{jt^*}$$

where $t^* = t - \sum_{i'<i} D_{i'} - D_i/2$. The expected number of projects from the current portfolio that are still active in development at time *t*, in Phase i, can then be calculated as

$$M_{it}^K = \sum_j K_{ij0} \pi_{jt}^i$$

Similarly, the expected number of projects still active at time *t*, in disease area *j*, can be calculated as

$$E_{jt}^K = \sum_i K_{ij0} \pi_{jt}^i$$

The expected cost incurred by the current development portfolio at time t, is given by

$$\Gamma_t^K = \sum_j \sum_i K_{ij0} C_{ij} \pi_{jt}^i$$

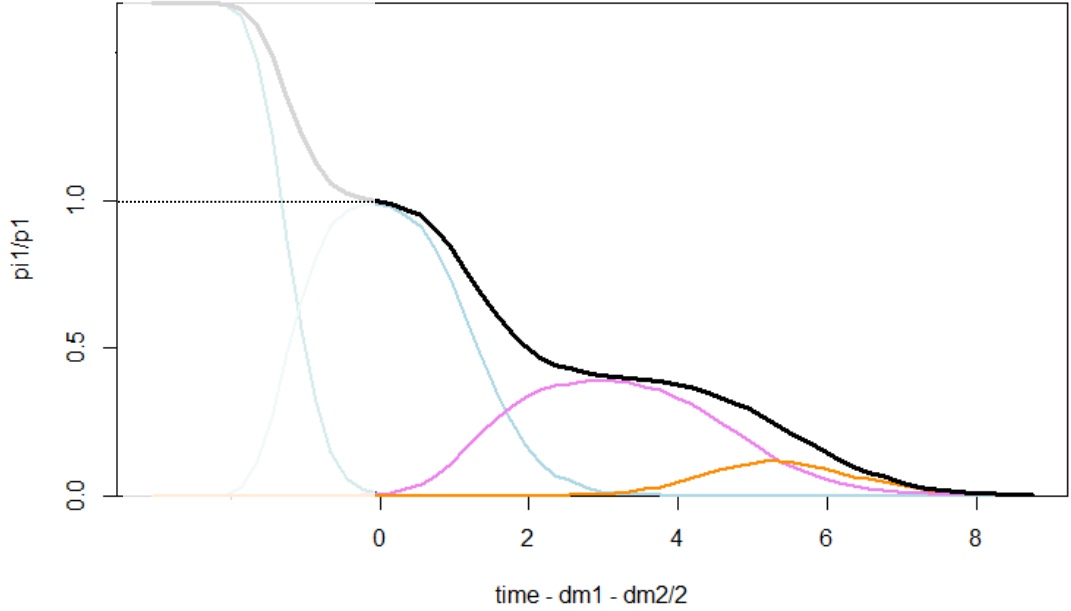

*Figure 4. An illustrating example of the probability, $\pi_{jt}^2$, (bold black line) that a project currently in Phase 2 remains active in development over time. The thinner colored lines to the right of 0 show the probabilities $\pi_{2t}^2, \pi_{3t}^2, \pi_{rt}^2$, that a project currently in Phase 2 will be active in Phase 2, 3, or registration, respectively (from 0 to right).*

## Future additions to the portfolio

A focus of this article is to assess the optimal addition to the company's portfolio over the coming planning period. Following the definitions above, a project that entered development at time $\tau$, will after $t$ years be in Phase $i$, with the probability given by $\pi_{ij,t-\tau}$. Let $N_{j\tau}$ denote the number of projects that will enter phase 1 of clinical development in disease area $j$, at time $\tau$. The expected number of projects active in Phase $i$ at time $t$ is then given by

$$M_{it}^N = \sum_j \sum_{\tau=0}^t N_{j\tau} \pi_{ij,t-\tau}$$

Similarly, the expected number of projects active in development in disease area $j$ at time $t$ is given by

$$E_{jt}^N = \sum_i \sum_{\tau=0}^t N_{j\tau} \pi_{ij,t-\tau}$$

The expected costs incurred at time t from future additions to the portfolio is

$$\Gamma_t^N = \sum_j \sum_{\tau=0}^t N_{j\tau} \Gamma_{j,t-\tau}$$

Combining the current portfolio with the future additions, we get $M_{it} = M_{it}^K + M_{it}^N$ as the risk-adjusted expected number of projects in Phase i at time $t$. Similarly, $E_{jt} = E_{jt}^K + E_{jt}^N$ is the risk-adjusted expected number of projects in disease area *j* at time *t*. The expected cost incurred by the development portfolio at time t is $\Gamma_t = \Gamma_t^K + \Gamma_t^N$.

The number of projects, $N_{j\tau}$, that will be brought into clinical development is a key parameter to achieve the long-term strategic goals of the company. It is a parameter that the company management can regulate by its investments and efforts in early research activities. This parameter is also a key input to the model that we develop, and it is the key decision variable that will go into the optimization procedure that will be presented in a later section of this paper.

## Number of launched products

We defined in previous paragraphs the duration of each phase, $D_{ij}$, and the probability of successful transition between phases, $P_{ij}$. Let $\Delta_j$ be the total duration of the development phases, $\Delta_j = \sum_{i=1}^{r} D_{ij}$, and let $Q_j$ be the aggregated probability of launch for a project, $Q_j = \prod_{i=1}^{r} P_{ij}$. With the input parameters given, the probability that a project which entered development at time $\tau$ is successfully launched to the market at time *t*, can be calculated as

$$\mu_{j,t-\tau} = Pr(\Delta_j < t - \tau) \cdot Q_j$$

i.e. the probability that all development phases have been completed in less than *t-τ* years, multiplied by the product of all phase success probabilities. With $N_{j\tau}$ projects entering development at time *τ*, the expected number of launches in disease area *j* during a planning period of *T* years can be estimated as

$$L_j^N = \sum_{\tau=0}^{T} N_{j\tau} \mu_{j,T-\tau}$$

The probability that a project which is currently in Phase *i* will eventually be launched to the market is $\rho_j^i = \prod_{i' \geq i} P_{i'j}$, and the expected number of projects from the current portfolio that will be launched are

$$L_j^K = \sum_{i=1}^{r} K_{ij0} \rho_j^i$$

The expected total number of launched projects in disease area *j* during the planning period is then $L_j = L_j^K + L_j^N$.

## Sales revenue

We will assume that there exist commercial forecasts regarding the anticipated sales for the products that the company currently has on the market. Similarly, the company has sales forecasts for the projects currently in development. We let $R_t^M$ denote the aggregated sales forecast for all marketed products and let $R_t^K$ denote the aggregated risk-adjusted forecasts for products in development.

Projects to be added to the development portfolio will be approximated to follow a simple ramp-up revenue model. The revenue gained from a project, after having been launched to the market is

$$R_{jt} = \begin{cases} \dfrac{t}{U} \cdot PYR_j & 0 < t \leq U \\ PYR_j & U < t \leq \Lambda \\ \lambda \cdot PYR_j & \Lambda < t \end{cases}$$

where $PYR_j$ is the peak year revenue from a project in disease area $j$, $U$ is the time it takes for sales to ramp-up to the peak, $\Lambda$ is the length of exclusivity on the market and $\lambda$ is the fraction of revenue that remains after loss of market exclusivity.

For a project that enters development at time $\tau$, and correspondingly being launched to the market at time $\tau + \Delta_j$, we have the revenue curve as $R_{j,t-\tau-\Delta_j}$. The expected revenue from the projects added to the portfolio over the planning period is then

$$R_t^N = \sum_j \sum_{\tau=0}^{T} N_{j\tau} Q_j R_{j,t-\tau-\Delta_j}$$

The total risk-adjusted sales forecast for the company is obtained by adding the revenue from the currently marketed products, future launches from the current development portfolio, and the future additions to the portfolio, i.e. $R_t = R_t^M + R_t^K + R_t^N$.

# Optimization

## Framing the optimization problem

Strategic management of a portfolio of drug development projects should involve the planning for the future addition of projects required to achieve the targets defined for the portfolio. Figure 5 provides a schematical illustration, where the blue areas represent the revenue forecasts for products currently on the market and projects currently in development. The orange area represents the cost expected for the projects currently in development. Additionally, the graph shows the revenue target set for the company (blue dashed line) and the anticipated development budget (dashed red line). The problem facing the portfolio manager is to use the available room in the budget (yellow area) by adding new projects to the portfolio in the best way to fill the revenue target gap (green area). The appropriate addition of projects is achieved by the selection of the optimal number of projects to be added at each year, in each of the targeted disease areas. In our notation, the optimization problem is thus to select the optimal values of $N_{j\tau}$.

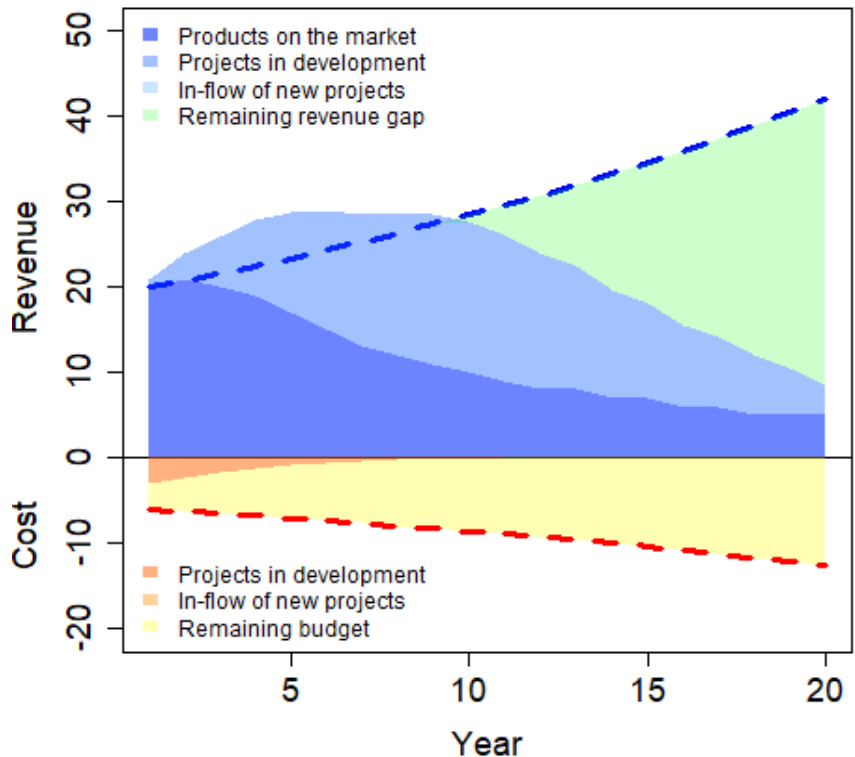

*Figure 5 Revenue and development cost cash flows from the current portfolio of marketed products and development projects.*

## Choice of objective functions and constraints

The optimization problem that is relevant to the decision-maker, i.e. the optimal choice of $N_{j\tau}$, could be defined differently depending on the focus of the strategic goal for the portfolio management. This was briefly outlined in Table 1 in the Introduction, and we will in the following paragraphs define the optimization problems in more detail. The description will to a large extent follow the structure outlined in Wiklund et al (2023).

Our quantitative model implies that the objective function to be optimized is related to a time series, i.e. the cost and revenue, respectively, are to be minimized/maximized over time. It is not obvious how to minimize/maximize these time series, and for each of the four strategic goals, we therefore provide two variants of the objective functions. One version of the objective function is focussing on the mean of the time series, and the other version is based on the min/max value of the time series over the planning horizon.

A number of constraints are added to the optimization problem. In order to achieve a healthy and diversified portfolio, the company may decide to maintain a portfolio that is expected to have at least $F_i$ development projects active in each phase, to have at least $G_j$ development projects active in each disease area, and to expect at least $H_j$ launches from each disease area over the planning period.

An additional constraint to the optimization problem is related to the capacity of the organisation in producing new projects for clinical development. This capacity is built gradually over time and cannot be expected to be varied extensively from year to year. To represent this inertia, we introduce a restriction on the annual increase in number of projects, $N_\tau \leq N_{\tau-1} + \delta$, where $\delta$ is the maximal annual increase in number of new projects.

### 1. Achieve a mean revenue target over time

One possible strategic goal for the portfolio is to achieve an acceptable mean revenue over the planning period. This implies choosing the timing and type of project inflow, i.e. choose the values of $N_{j\tau}$, so that the mean revenue over the planning horizon reaches a given target, $S$, while keeping the total development costs as low as possible. Stated as a formal optimization problem, we have $N_{j\tau}$ as the decision variables, and we want to

$$\underset{N_{j\tau}}{\text{minimize}} \sum_{t=1}^{T} \Gamma_t \tag{1A}$$

$$\text{subject to: } \left(T^{-1} \sum_{t=1}^{T} R_t\right) \geq S$$

$$M_{it} \geq F_i$$
$$E_{jt} \geq G_j$$
$$L_j \geq H_j$$
$$N_\tau \leq N_{\tau-1} + \delta$$

The objective function above would allow a volatile cost and could lead to very large costs in some years. An alternative would be to select an objective function that restricts the peak cost, i.e.

$$\underset{N_{j\tau}}{\text{minimize}} \left(\max_t (\Gamma_t - B_t)\right) \tag{1B}$$

This objective function implies that we want to ensure that the amount by which the cost in any year exceeds the budget, should be kept as small as possible.

### 2. Achieve a target revenue each year

With this strategic goal, we prioritize a stable revenue flow and choose the values of $N_{j\tau}$ so that the revenue in each year reaches a given target, $S_t$, while keeping the total costs as low as possible. Stated as a formal optimization problem, we want to

$$\underset{N_{j\tau}}{\text{minimize}} \sum_{t=1}^{T} \Gamma_t \tag{2A}$$

$$\text{subject to: } R_t \geq S_t. \quad \text{for all } t \in \{1,..,T\}$$
$$M_{it} \geq F_i$$
$$E_{jt} \geq G_j$$
$$L_j \geq H_j$$
$$N_\tau \leq N_{\tau-1} + \delta$$

As in the previous paragraph, the objective function above would allow a volatile cost structure and an alternative objective function would be to select an objective function that restricts the peak cost as compared to budget, i.e. we want to

$$\underset{N_{j\tau}}{\text{minimize}} \left(\max_t (\Gamma_t - B_t)\right) \tag{2B}$$

### 3. Keep mean development cost within budget over time

With this strategic goal, we want to ensure that the mean development cost does not exceed the assigned total budget over the planning horizon. The values of $N_{j\tau}$ are chosen so that the mean cost is contained within a given budget, $B$, while reaching a revenue as high as possible. Stated as a formal optimization problem, the objective is to

$$\underset{N_{j\tau}}{\text{maximize}} \sum_{t=1}^{T} R_t \tag{3A}$$

$$\text{subject to:} \quad \left( T^{-1} \sum_{t=1}^{T} \Gamma_t \right) \leq B$$
$$M_{it} \geq F_i$$
$$E_{jt} \geq G_j$$
$$L_j \geq H_j$$
$$N_\tau \leq N_{\tau-1} + \delta$$

The objective function above could lead to volatile revenue flow, and an alternative would be to select an objective function that ensures that a potential revenue deficit as compared to the revenue target is minimized

$$\underset{N_{j\tau}}{\text{maximize}} \left( \underset{t}{\min}(R_t - S_t) \right) \tag{3B}$$

### 4. Keep development cost within budget each year

With this strategic goal, a stable cost flow is prioritized. We choose the values of $N_{j\tau}$ so that the cost in any year does not exceed the budget, $B_t$, while achieving a total revenue that is high as possible. Stated as a formal optimization problem, we want to

$$\underset{N_{j\tau}}{\text{maximize}} \sum_{t=1}^{T} R_t \tag{4A}$$

$$\text{subject to:} \quad \Gamma_t \leq B_t. \quad \text{for all } t \in \{1,..,T\}$$
$$M_{it} \geq F_i$$
$$E_{jt} \geq G_j$$
$$L_j \geq H_j$$
$$N_\tau \leq N_{\tau-1} + \delta$$

As in the previous paragraph, the objective function above would allow a volatile revenue flow. An alternative would be to select an objective function that keeps potential revenue deficits as small as possible, i.e. we want to

$$\underset{N_{j\tau}}{\text{maximize}} \left( \underset{t}{\min}(R_t - S_t) \right) \tag{4B}$$

## Optimization methods

Optimization was done with a simulated annealing algorithm (e.g., Givens & Hoeting, 2013, Section 3.3) for combinatorial optimization. Starting with an arbitrary design which fulfills all constraints, the design is iteratively improved. In each iteration, a candidate design is generated by adding one project and/or removing one project randomly but requiring that it still fulfils all constraints. The candidate design is accepted with probability according to the simulated annealing rule. This means that candidates which improve the criterion value are always accepted, and candidates which deteriorate the criterion value are accepted sometimes where the acceptance probability reduces with larger iteration number and with larger discrepancy in the criterion values between last design and candidate. The hyperparameters in the algorithm, like temperature schedule and iteration number were chosen after running several cases to ensure that the solution is close to optimal.

# Illustrating example

## Input assumptions

We will in this section present the use of the proposed methodology and the model developed in previous sections by means of an illustrating example. The example considers a company which is looking to achieve a desired revenue increase over an extended planning period. The forecasts for the company's revenue from marketed products and from assets in the current development portfolio are illustrated in Figure 5, together with the forecasted cost incurred by the current development projects. Figure 5 also shows the desired revenue target and budget limitations. It illustrates that the company's forecasted revenue is expected to drop as the sales of marketed drugs will decrease, leaving a substantial gap to be filled to achieve the desired revenue target. The company will now explore strategies of adding new projects to the development portfolio, with respect to the different ways of framing objectives for the portfolio optimization outlined in Table 1.

The company aspires to have a balanced and diverse development portfolio. This implies having active projects in all phases of development and having a presence across the disease areas. They also aim for a healthy level of market launches during the planning period. These requirements for the portfolio are quantified in the parameters, $F_i$, $G_j$ and $H_j$, as specified in Table 2. The composition of the company's current portfolio is given in Table 3.

*Table 2. The constraints defined for the portfolio optimization, in terms of least number of projects in each phase, least number of projects in each disease area, and least number of launched projects.*

|         | Min #projects in Phase |                | Min #projects in disease area | Min #projects launched |
|---------|------------------------|----------------|-------------------------------|------------------------|
| Phase 1 | $F_1$=12               | Disease area 1 | $G_1$=10                      | $H_1$=12               |
| Phase 2 | $F_2$=6                | Disease area 2 | $G_2$=5                       | $H_2$=5                |
| Phase 3 | $F_3$=3                | Disease area 3 | $G_3$=8                       | $H_3$=10               |

*Table 3. The number of projects, $K_{ij0}$, in the company's current development portfolio, in each phase of development and in each disease area.*

|              | Disease area 1 | Disease area 2 | Disease area 3 |
|--------------|----------------|----------------|----------------|
| Phase 1      | 6              | 2              | 5              |
| Phase 2      | 3              | 2              | 2              |
| Phase 3      | 3              | 1              | 1              |
| Registration | 1              | 1              | 1              |

The key input parameters of the project model, regarding cost, duration and success probabilities for the three disease areas are given in Table 4. The costs and durations are unknown at the time of modelling, and to represent this uncertainty we expect these parameters to be given as a stochastic distribution. We assign a lognormal distribution to these parameters, as $D_{ij} \sim lognormal(ln(\delta_{ij}), \sigma_j)$ and $C_{ij} \sim lognormal(ln(\gamma_{ij}), \sigma_j)$. The median duration, $\delta_{ij}$, and median cost, $\gamma_{ij}$, are the parameters given in Table 4. The aggregated sales forecasts for the products currently on the market are given in Table 5, and sales forecast parameters for development projects are given in Table 6.

Table 4. Key input parameters of the project model used for the illustrating example.

|  |  | Phase 1 | Phase 2 | Phase 3 | Registration |  |
|---|---|---|---|---|---|---|
|  |  | Median duration in each phase, $\delta_{ij}$ (years) |  |  |  | Variability, $\sigma_j$ |
| Duration | Disease area 1 | 2 | 2.5 | 3.5 | 1 | 0.4 |
|  | Disease area 2 | 2 | 3 | 4 | 1.5 | 0.4 |
|  | Disease area 3 | 1.5 | 2 | 3 | 1 | 0.4 |
|  |  | Median cost in each phase, $\gamma_{ij}$ ($Bn) |  |  |  | Variability, $\sigma_j$ |
| Cost | Disease area 1 | 0.2 | 0.3 | 0.6 | 0.1 | 0.4 |
|  | Disease area 2 | 0.15 | 0.3 | 0.5 | 0.1 | 0.4 |
|  | Disease area 3 | 0.2 | 0.4 | 0.8 | 0.1 | 0.4 |
|  |  | Probability of transition to next phase. $P_{ij}$ |  |  |  | Prob. f launch, $Q_j$ |
| Success probability | Disease area 1 | 0.6 | 0.4 | 0.7 | 0.95 | 0.16 |
|  | Disease area 2 | 0.5 | 0.3 | 0.6 | 0.9 | 0.08 |
|  | Disease area 3 | 0.6 | 0.3 | 0.6 | 0.9 | 0.11 |

Table 5. Sales forecasts over the planning period, for projects currently on the market ($Bn).

| Year | Forecasted revenue, $R_t^M$ ($Bn) |
|---|---|
| 1 | 20 |
| 2 | 21 |
| 3 | 21 |
| 4 | 20 |
| 5 | 18 |
| 6 | 16 |
| 7 | 15 |
| 8 | 14 |
| 9 | 13 |
| 10 | 12 |
| 11 | 11 |
| 12 | 10 |
| 13 | 10 |
| 14 | 9 |
| 15 | 9 |
| 16 | 8 |
| 17 | 8 |
| 18 | 7 |
| 19 | 7 |
| 20 | 6 |
| 21 | 6 |
| 22 | 5 |

| 23 | 5 |
|---|---|
| 24 | 5 |
| 25 | 4 |
| 26 | 4 |
| 27 | 4 |
| 28 | 4 |
| 29 | 3 |
| 30 | 3 |

*Table 6. Sales forecast parameters for a development project in each of the disease areas.*

|  | Ramp up, $U$ (years) | Peak revenue, $PYR_j$ ($Bn) | Fraction of sales after LOE, $\lambda$ ($Bn) |
|---|---|---|---|
| **Disease area 1** | 3 | 1.5 | 0.2 |
| **Disease area 2** | 5 | 5 | 0.1 |
| **Disease area 3** | 4 | 3 | 0.15 |

In each iteration of a Monte Carlo simulation, values are drawn from the distributions to populate input parameters. Given these values, key characteristics of the project model are calculated for each iteration, and the mean of values over the iterations provide risk-adjusted averages of key characteristics. For example, key model characteristics such as the probability that a project in disease area $j$ is active in development at time $t$, $\pi_{jt}$, the risk-adjusted costs $\Gamma_t$ and the launch probabilities $\mu_{jt}$, are calculated as averages from the simulations. These risk-adjusted measures are fed into the optimization procedure for the planning of future additions to the portfolio. For given values of $N_{j\tau}$ we can then calculate the key output metrics $E_{jt}$, $M_j$ and $R_{jt}$. Through the optimization procedure we may obtain the values of $N_{j\tau}$ that are optimal with respect to the chosen optimization criteria.

In the calculations for the illustrating examples, we are using a time horizon of T=30 years. Projects entering development in the latter part of that period will not generate any revenue within the time horizon, and this will have an impact on the optimal solution when it comes to inflow of projects in the last years. For this reason, following the convention of Wiklund et al (2023), we will only use results for the first 10 years when showing the optimal number of added projects, and graphs illustrating the cost and revenue are restricted to the first 20 years.

## Results

Results from the illustrating example are provided in Figures 6-29. For each of the eight ways of framing the company's strategic goal, leading to optimization problems 1A-4B as described above, the results are summarized in three figures; (i) a bar chart showing the optimal solution of the number and timing of the addition of new projects in each disease are, (ii) a graph showing the development of expected revenue and cost over time for the optimal solution, (iii) line graphs showing the development over time with relation to the restrictions regarding expected number of projects in each phase and expected number of projects in each disease area.

For the strategic goal 1A (Fig. 6-8), the objective is to achieve a mean revenue target over time while minimizing the total development costs. The optimal strategy is then to maximize initial inflow of projects, focusing on the disease areas 2 and 3, followed by a reduced inflow in a later part of the planning period. This strategy, however, leads to substantially exceeding budget in the first years. We see further that for disease area 1, the expected number of projects is close to the boundary in years 4 to 10 and the expected number of projects launched is also the minimum required (Table 2). Therefore, the inflow of project in this area are to meet these constraints.

With strategic goal 1B (Fig. 9-11), the objective is to achieve a mean revenue target, while ensuring that the cost, relative to the designated budget, should be kept as small as possible. With this objective, the strategy is to focus primarily on projects in disease area 3, keeping the number of new projects relatively constant. While this strategy keeps expected costs within budget, it is expected not to completely achieve target revenue in some parts of the planning period.

With strategic goal 2A (Fig. 12-14), a target revenue should be achieved each year, while keeping the total cost over the planning period as low as possible. For this strategy, an initial focus on projects from disease area 3 should be shifted to prioritizing inflow of projects from disease area 2 in a later part of the planning period. This strategy is expected to exceed the budget in parts of the period.

Strategic goal 2B (Fig. 15-17) is the case where a target revenue should be achieved each year, while ensuring that the cost, relative to the designated budget, should be kept as small as possible. With this strategy, the expected revenue target is achieved while keeping the budget for most of the period. It is achieved with a relatively constant inflow of new projects, with a slight focus on projects in disease area 3.

For the strategic goal 3A (Fig. 18-20), the objective is that total cost over the planning period should be within the assigned budget, and total revenue should be maximized given that restriction. Since both the total budget can flexibly be spend during the 30-year period and the time when the revenues are earned does not matter for this objective, the optimal strategy is to maximize the inflow of project in the early part of the planning period, gradually focusing mainly on disease area 2. This strategy leads to a substantial excess over budget in large parts of the planning period. In reality, it is not realistic to make these substantial investments over budget and we see therefore that the full flexibility of goal 3A can generally lead to non-feasible results.

With strategic goal 3B (Fig. 21-23), the objective is to keep the total cost over the period within budget, and any deficit to revenue target should be kept as small as possible. This leads to a strategy with a relatively constant inflow of new projects, and a fairly equal distribution over the three disease areas. The strategy implies that costs moderately exceed budget for substantial parts of the period.

With strategic goal 4A (Fig. 24-26), the development costs are required to be within budget each year, and total revenue should be maximized given that restriction. This strategy is not expected to achieve the revenue target as the revenue from current development projects starts to decline. With this strategy, sourcing of new projects should be focused on disease area 2, keeping inflow from disease area 1 and 3 at a minimum to meet the constraints. disease area 2 is the area which has lower costs than the other areas and gives higher expected return on investment, but projects have longer expected development time. Therefore, when late revenues are equally beneficial as early revenues, disease area 2 projects are a good choice.

Strategic goal 4B (Fig. 27-29) is the case where the costs are required to be within budget, and any deficit to revenue target should be kept as small as possible. This leads to a strategy that is relatively similar to case 4A, with primarily a focus on new projects from disease area 2, while not completely reaching the revenue target for the entire period.

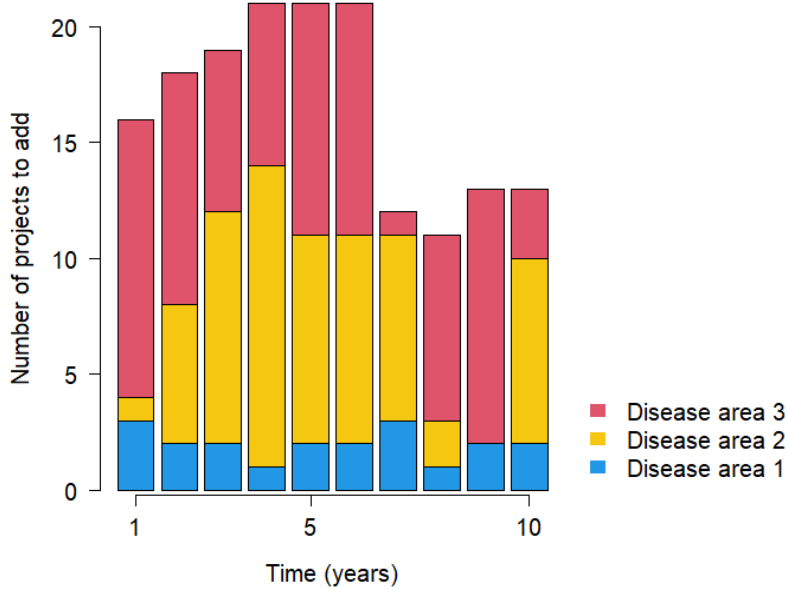

*Figure 6. The optimal number of added projects over time, in each disease area, for the optimization problem 1A. The goal is to achieve a mean revenue target over time, while minimizing the total development costs.*

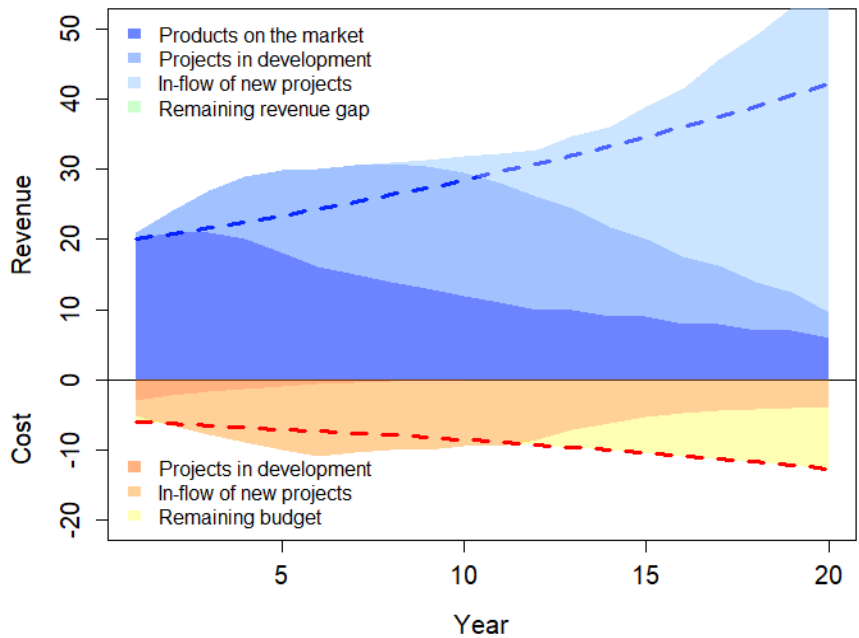

*Figure 7. Expected revenue and cost over time, given the optimal number of added projects, $N_{j\tau}$, for the optimization problem 1A. The goal is to achieve a mean revenue target over time, while minimizing the total development costs.*

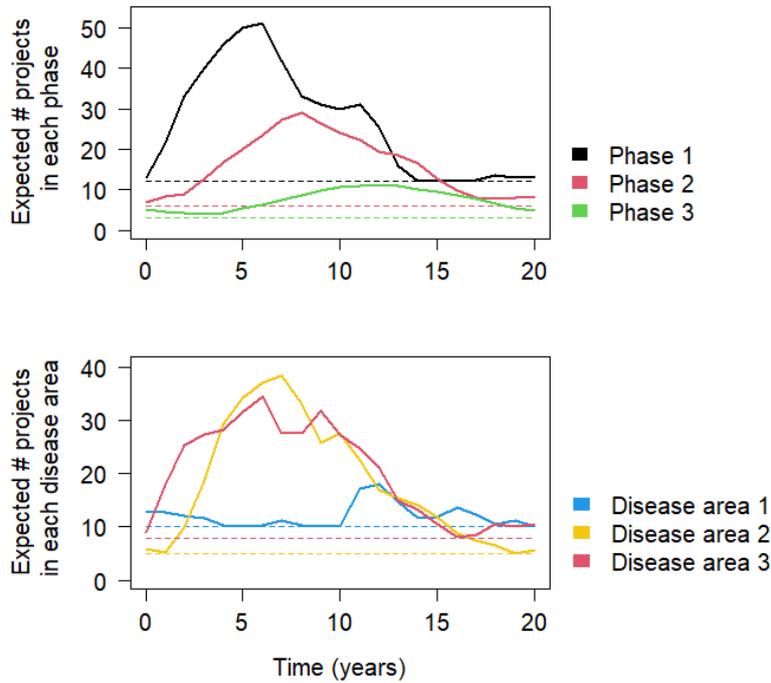

Figure 8. Expected number of projects in each phase and disease area over time, given the optimal number of added projects, $N_{j\tau}$, for the optimization problem 1A. Minimum constraints are indicated by dotted lines. The goal is to achieve a mean revenue target over time, while minimizing the total development costs.

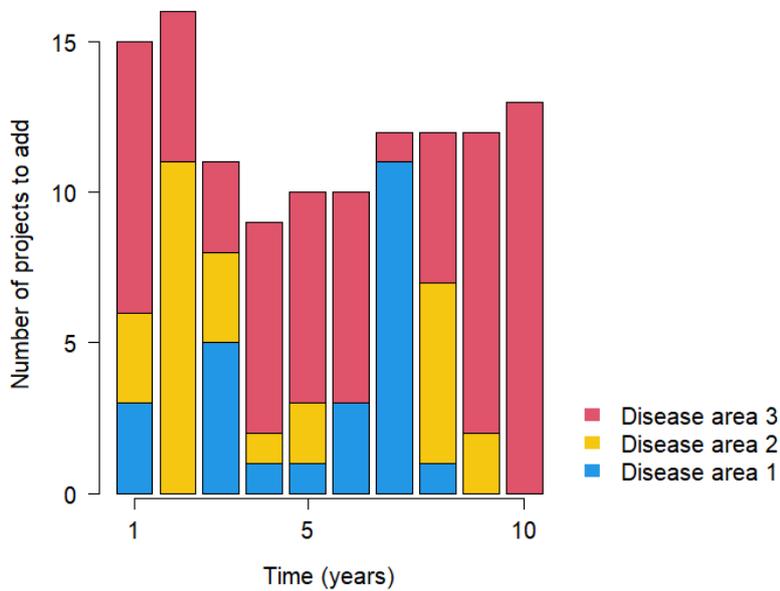

Figure 9. The optimal number of added projects over time, in each disease area, for the optimization problem 1B. The goal is to achieve a mean revenue target, while ensuring that the cost, relative to the allotted budget, should be kept as small as possible.

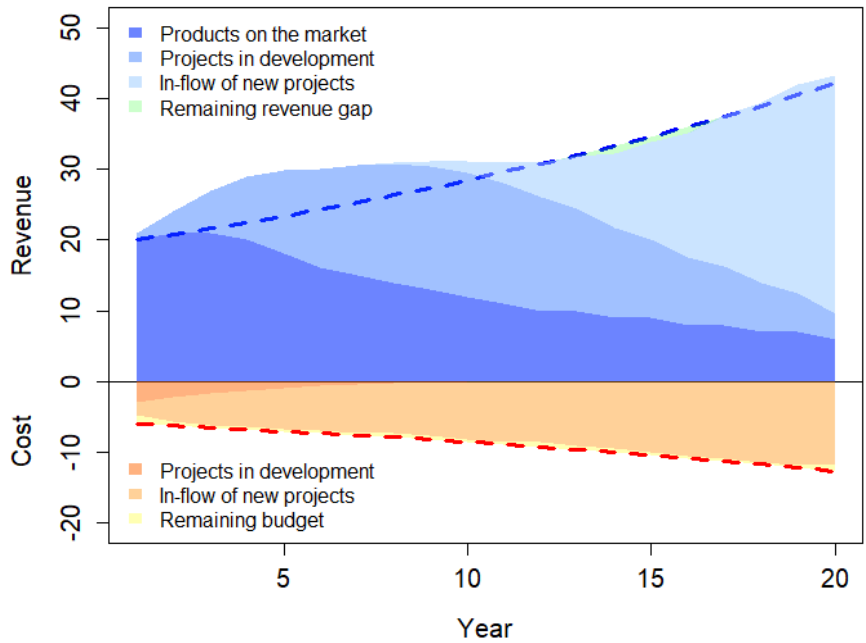

*Figure 10. Expected revenue and cost over time, given the optimal number of added projects, $N_{j\tau}$, for the optimization problem 1B. The goal is to achieve a mean revenue target, while ensuring that the cost, relative to the allotted budget, should be kept as small as possible.*

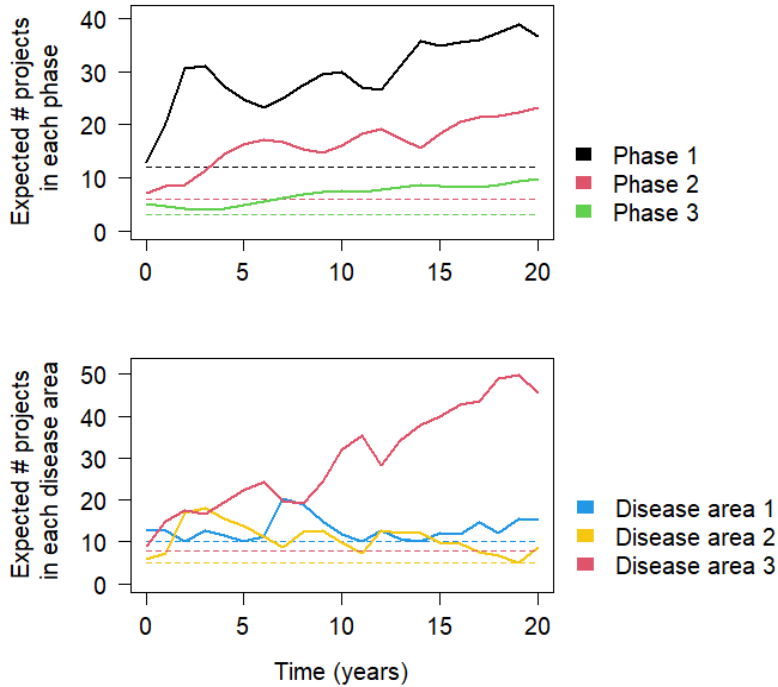

*Figure 11. Expected number of projects in each phase and disease area over time, given the optimal number of added projects, $N_{j\tau}$, for the optimization problem 1B. Minimum constraints are indicated by dotted lines. The goal is to achieve a mean revenue target, while ensuring that the cost, relative to the allotted budget, should be kept as small as possible.*

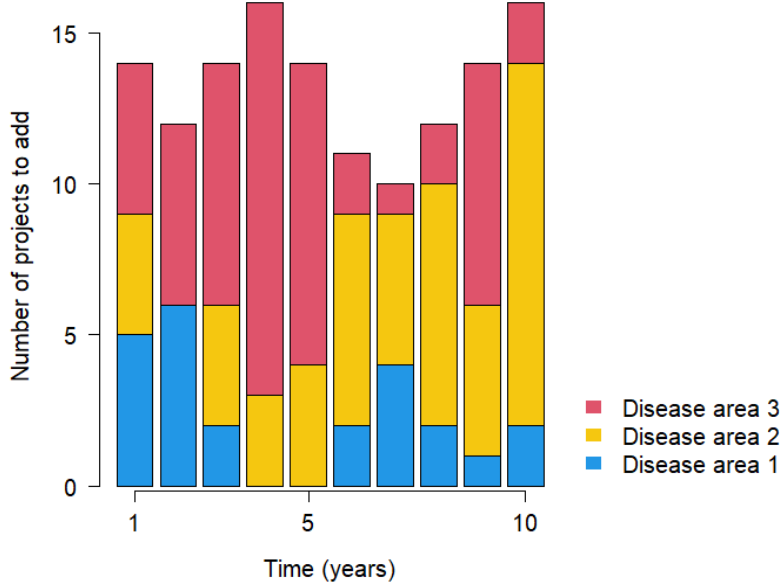

*Figure 12. The optimal number of added projects over time, in each disease area, for the optimization problem 2A. The goal is to achieve a target revenue each year, while keeping the total cost over the planning period as low as possible.*

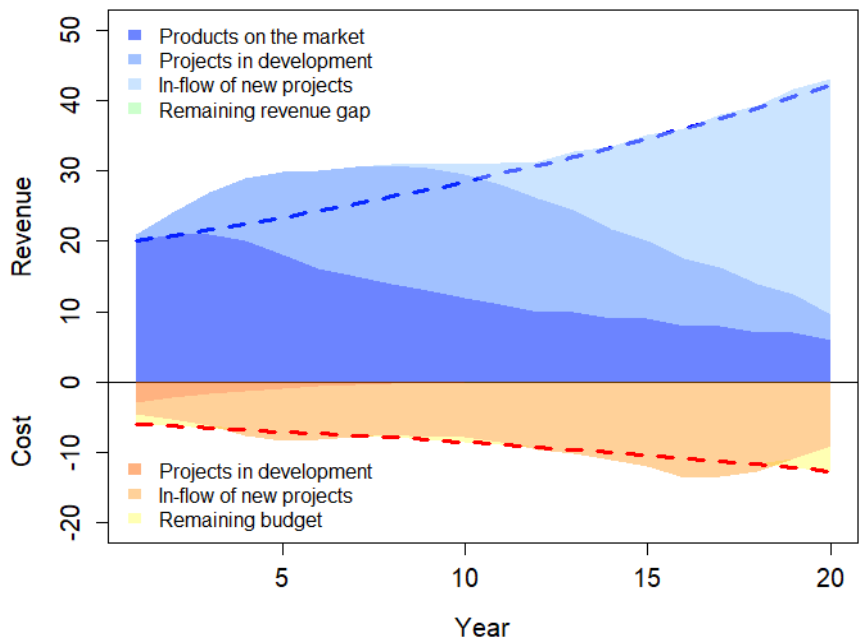

*Figure 13. Expected revenue and cost over time, given the optimal number of added projects, $N_{j\tau}$, for the optimization problem 2A. The goal is to achieve a target revenue each year, while keeping the total cost over the planning period as low as possible.*

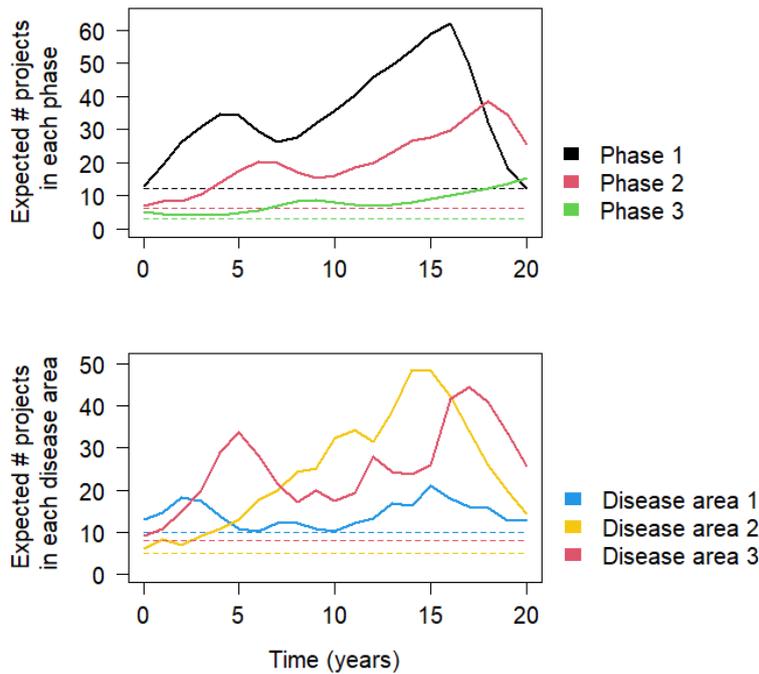

Figure 14. Expected number of projects in each phase and disease area over time, given the optimal number of added projects, $N_{j\tau}$, for the optimization problem 2A. Minimum constraints are indicated by dotted lines. The goal is to achieve a target revenue each year, while keeping the total cost over the planning period as low as possible.

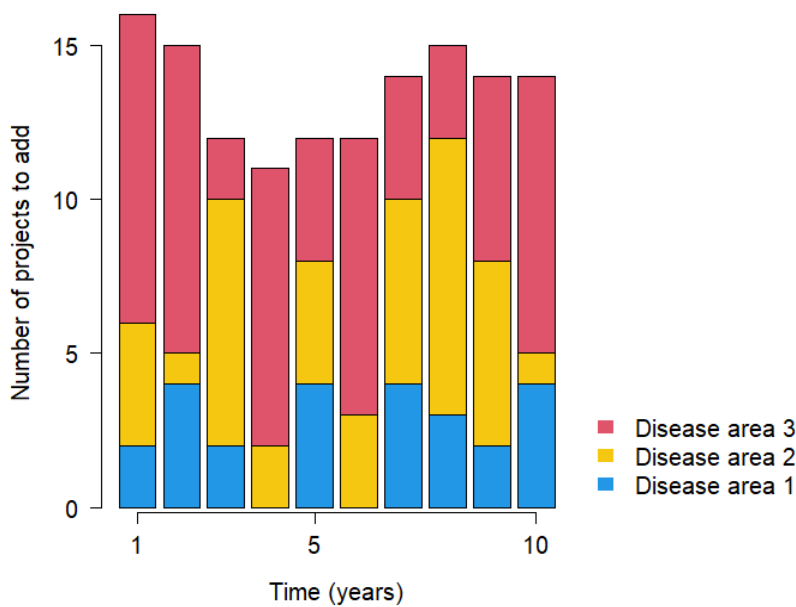

Figure 15. The optimal number of added projects over time, in each disease area, for the optimization problem 2B. The goal is to achieve a target revenue each year, while ensuring that the cost, relative to the allotted budget, should be kept as small as possible.

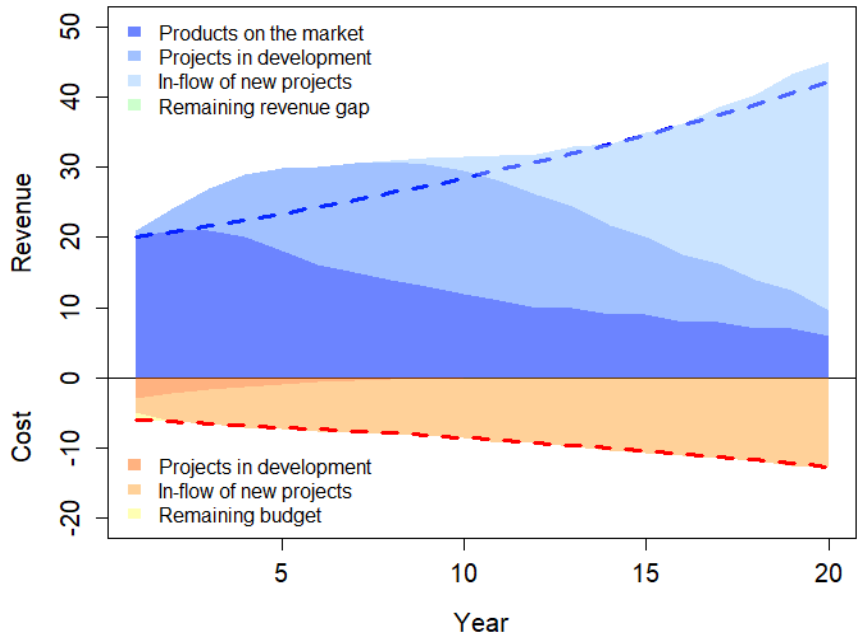

*Figure 16. Expected revenue and cost over time, given the optimal number of added projects, $N_{j\tau}$, for the optimization problem 2B. The goal is to achieve a target revenue each year, while ensuring that the cost, relative to the allotted budget, should be kept as small as possible.*

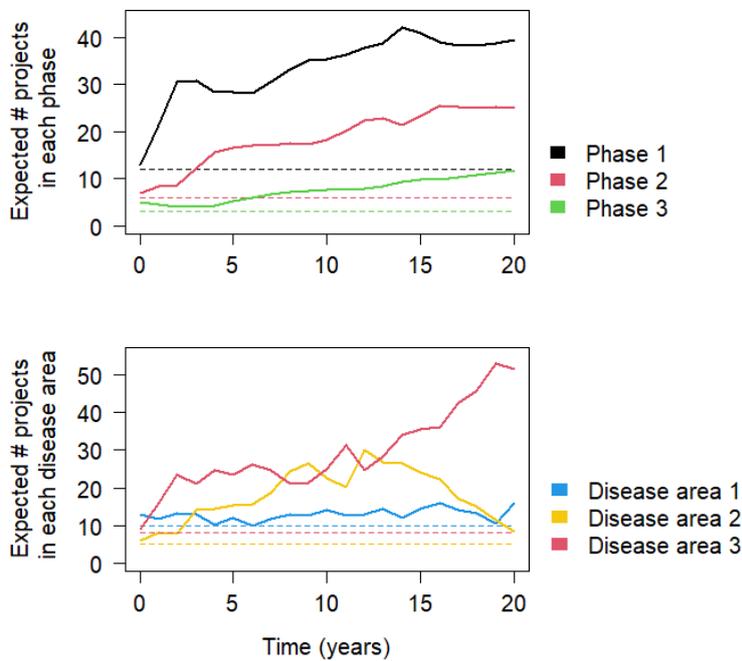

*Figure 17. Expected number of projects in each phase and disease area over time, given the optimal number of added projects, $N_{j\tau}$, for the optimization problem 2B. Minimum constraints are indicated by dotted lines. The goal is to achieve a target revenue each year, while ensuring that the cost, relative to the allotted budget, should be kept as small as possible.*

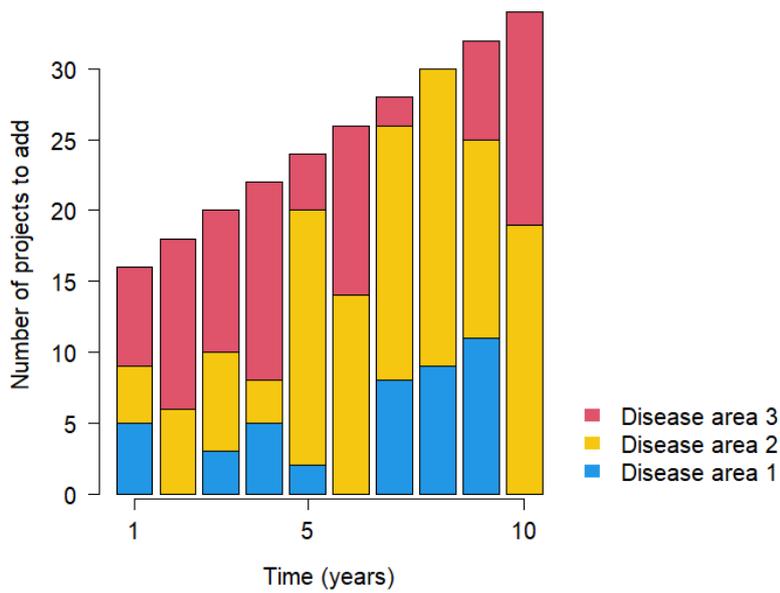

*Figure 18. The optimal number of added projects over time, in each disease area, for the optimization problem 3A. The goal is to keep the total cost over the planning period within the assigned budget, and total revenue should be maximized given that restriction.*

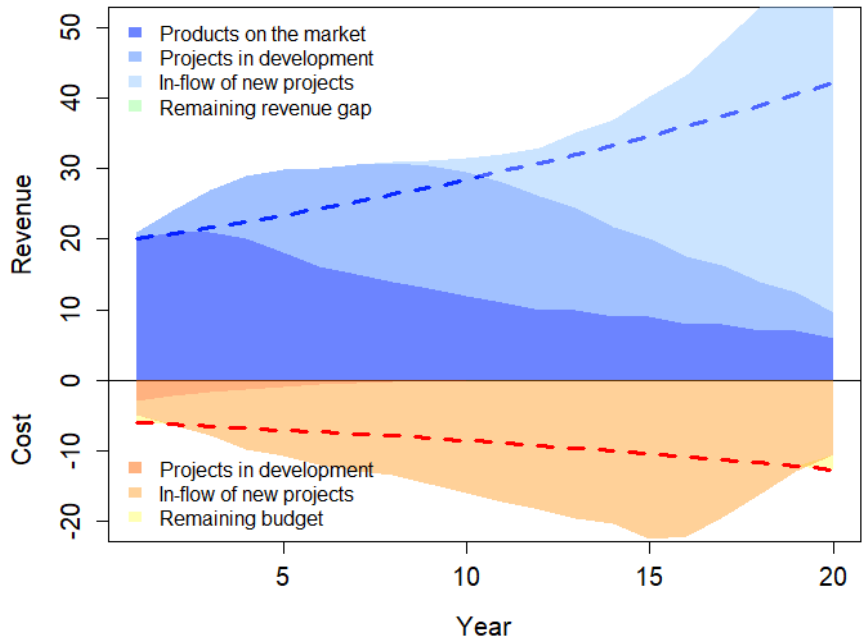

*Figure 19. Expected revenue and cost over time, given the optimal number of added projects, $N_{j\tau}$, for the optimization problem 3A. The goal is to keep the total cost over the planning period within the assigned budget, and total revenue should be maximized given that restriction.*

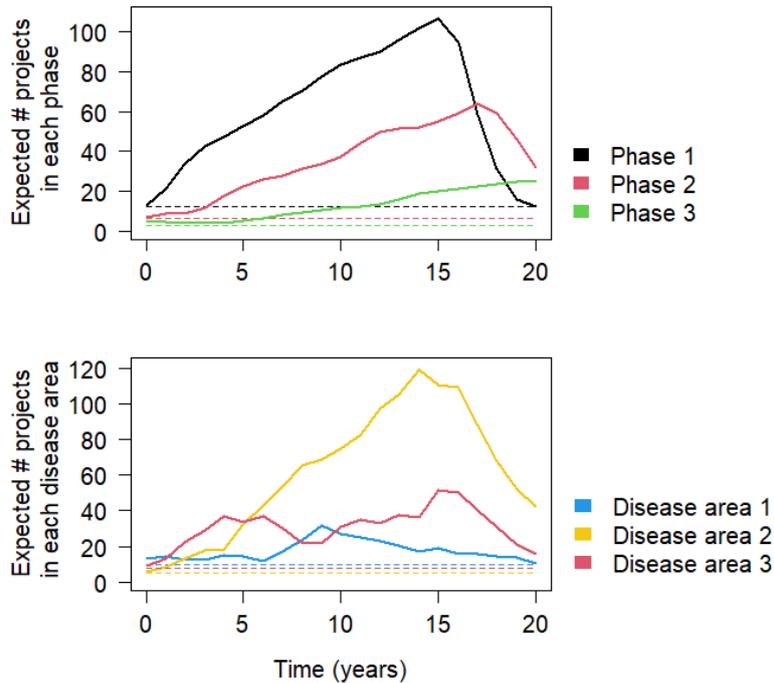

Figure 20. Expected number of projects in each phase and disease area over time, given the optimal number of added projects, $N_{j\tau}$, for the optimization problem 3A. Minimum constraints are indicated by dotted lines. The goal is to keep the total cost over the planning period within the assigned budget, and total revenue should be maximized given that restriction.

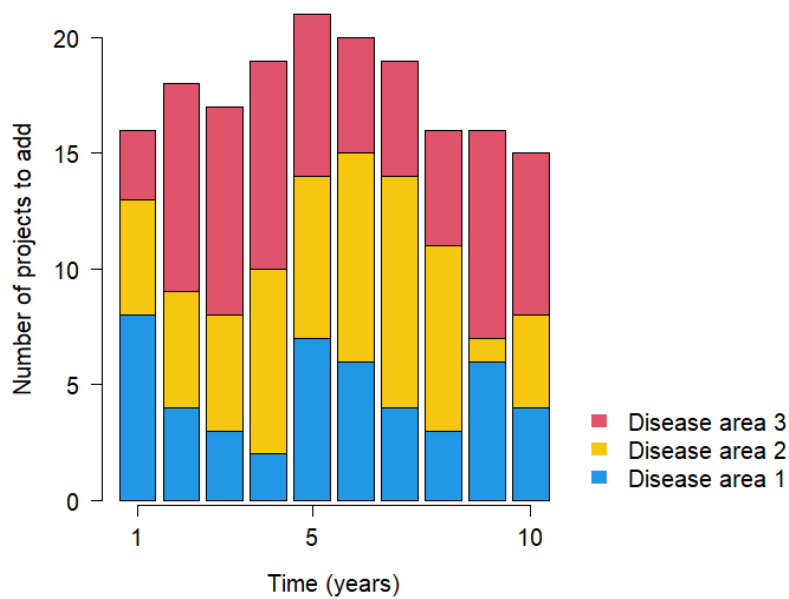

Figure 21. The optimal number of added projects over time, in each disease area, for the optimization problem 3B. The goal is to keep the total cost over the planning period within the assigned budget, and any deficit to revenue target should be kept as small as possible.

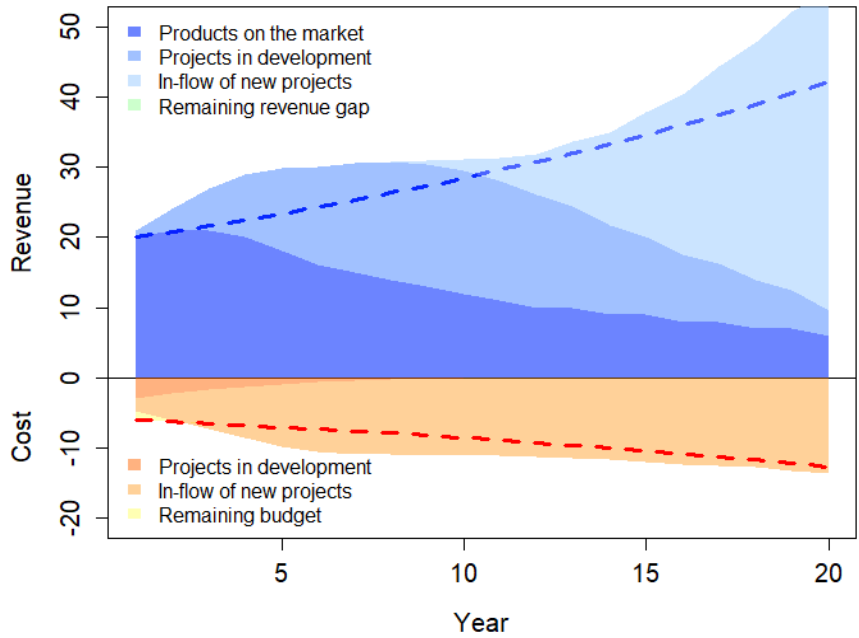

*Figure 22. Expected revenue and cost over time, given the optimal number of added projects, $N_{j\tau}$, for the optimization problem 3B. The goal is to keep the total cost over the planning period within the assigned budget, and any deficit to revenue target should be kept as small as possible.*

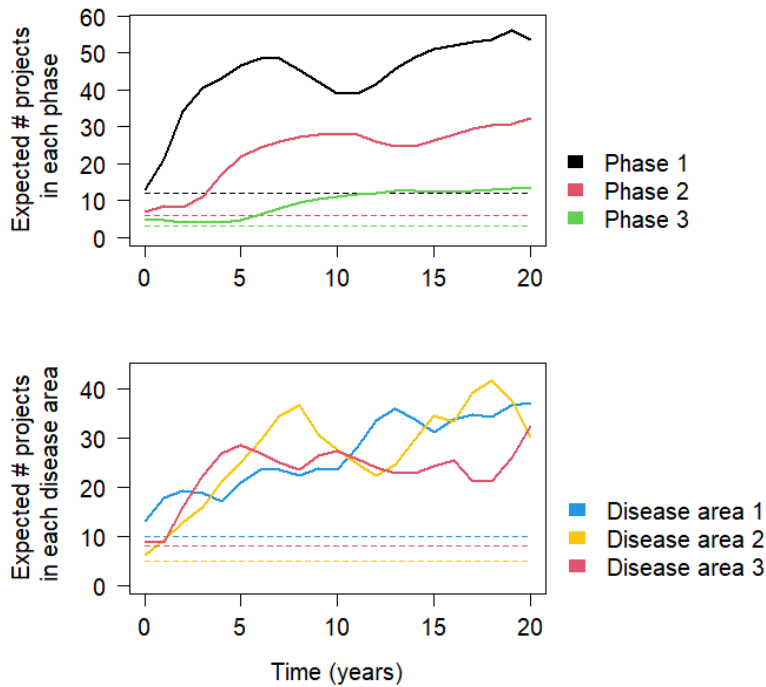

*Figure 23. Expected number of projects in each phase and disease area over time, given the optimal number of added projects, $N_{j\tau}$, for the optimization problem 3B. Minimum constraints are indicated by dotted lines. The goal is to keep the total cost over the planning period within the assigned budget, and any deficit to revenue target should be kept as small as possible.*

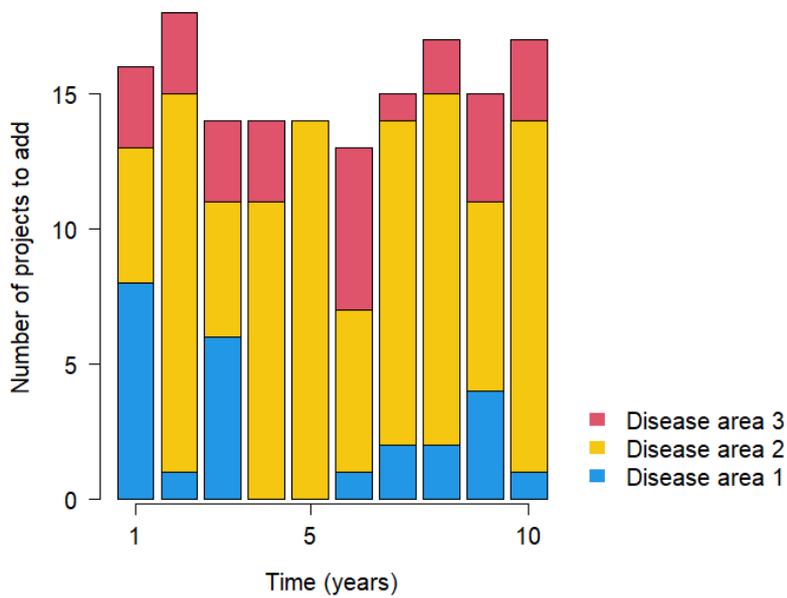

*Figure 24. The optimal number of added projects over time, in each disease area, for the optimization problem 4A. The goal is to keep the development costs within budget each year, and total revenue should be maximized given that restriction.*

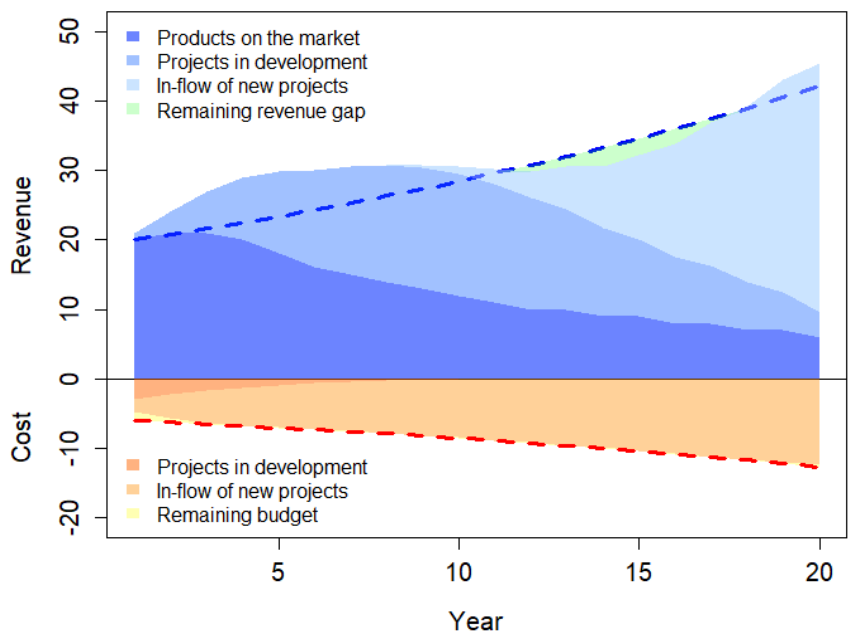

*Figure 25. Expected revenue and cost over time, given the optimal number of added projects, $N_{j\tau}$, for the optimization problem 4A. The goal is to keep the development costs within budget each year, and total revenue should be maximized given that restriction.*

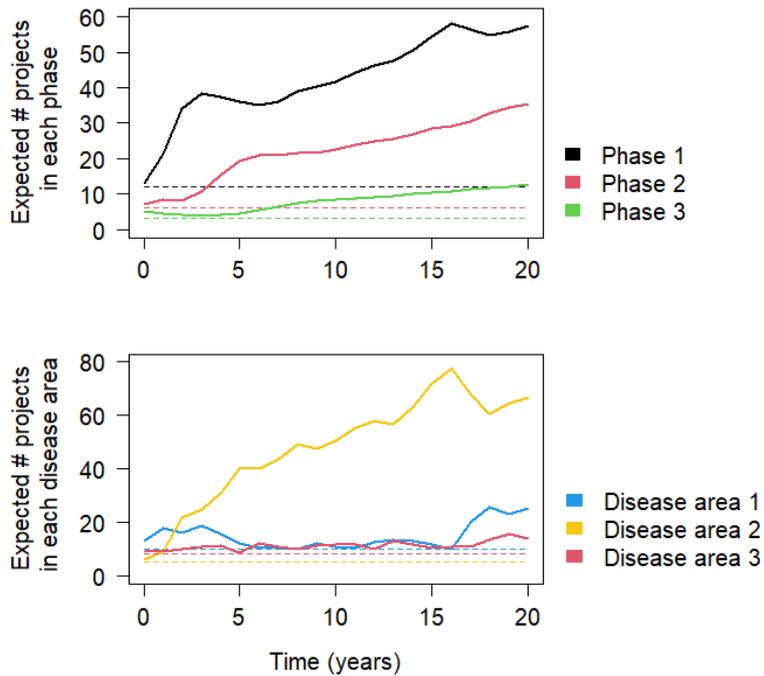

*Figure 26. Expected number of projects in each phase and disease area over time, given the optimal number of added projects, $N_{j\tau}$, for the optimization problem 4A. Minimum constraints are indicated by dotted lines. The goal is to keep the development costs within budget each year, and total revenue should be maximized given that restriction.*

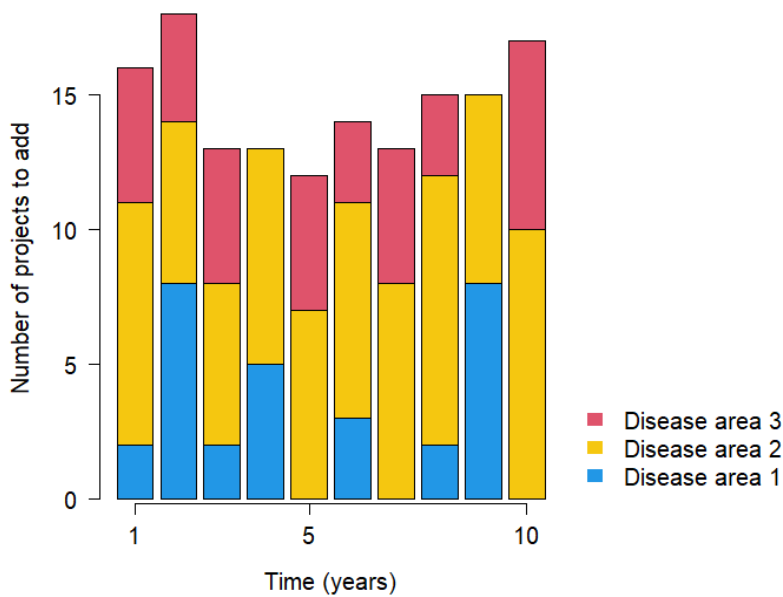

*Figure 27. The optimal number of added projects over time, in each disease area, for the optimization problem 4B. The goal is to keep the development costs within budget each year, and any deficit to revenue target should be kept as small as possible.*

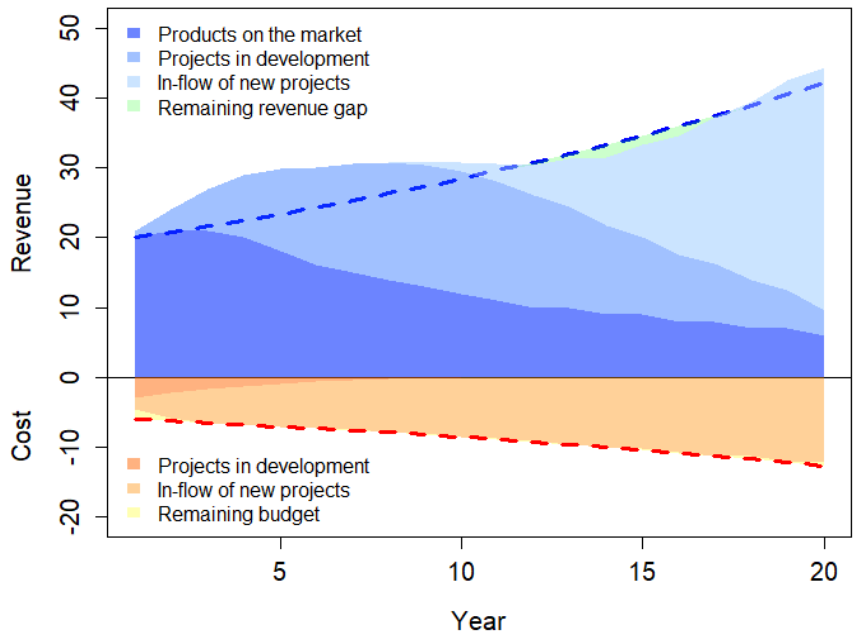

*Figure 28. Expected revenue and cost over time, given the optimal number of added projects, $N_{j\tau}$, for the optimization problem 4B. The goal is to keep the development costs within budget each year, and any deficit to revenue target should be kept as small as possible.*

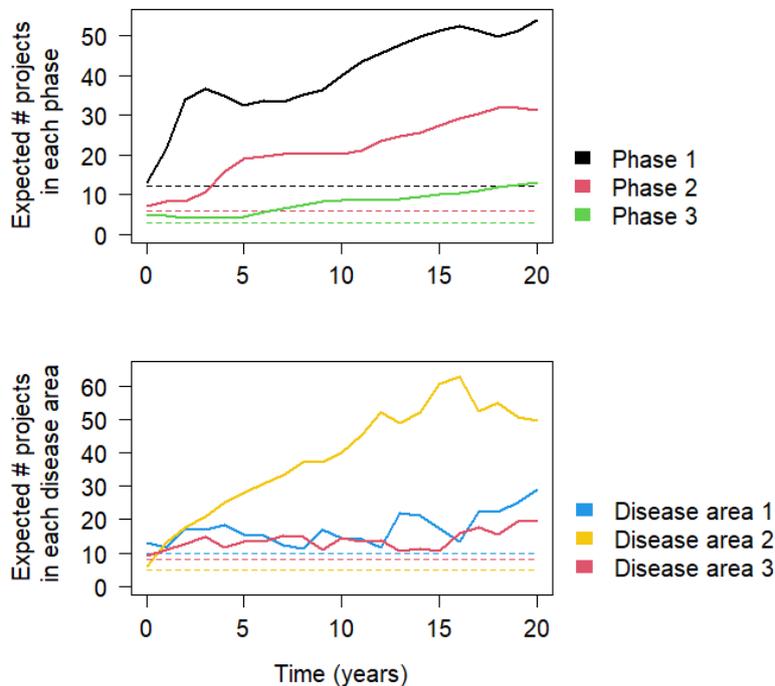

*Figure 29. Expected number of projects in each phase and disease area over time, given the optimal number of added projects, $N_{j\tau}$, for the optimization problem 4B. Minimum constraints are indicated by dotted lines. The goal is to keep the development costs within budget each year, and any deficit to revenue target should be kept as small as possible.*

*Table 7. Expected number of projects launched in each of the disease areas.*

|              | Disease Area 1 | Disease Area 2 | Disease Area 3 |
|--------------|----------------|----------------|----------------|
| **Case 1A**  | 12.0           | 8.3            | 15.6           |
| **Case 1B**  | 12.8           | 8.1            | 23.1           |
| **Case 2A**  | 13.8           | 8.9            | 21.3           |
| **Case 2B**  | 13.7           | 7.5            | 25.7           |
| **Case 3A**  | 13.4           | 25.0           | 22.7           |
| **Case 3B**  | 24.7           | 13.3           | 18.7           |
| **Case 4A**  | 12.9           | 18.4           | 16.7           |
| **Case 4B**  | 13.3           | 19.4           | 12.1           |
| **Min. required** | 12        | 5              | 10             |

# Discussion

In this article, we have shifted the focus from the traditional project selection problem to addressing the inflow of new projects into the pipeline. This approach significantly alters the strategic outlook of discrete R&D portfolio optimization. Traditional methods often produce solutions that are only momentarily optimal, requiring frequent reassessment as the portfolio evolves. Our approach, on the other hand, supports long-term strategic decision-making, which is crucial given the protracted timelines and high risk of failure in drug development. By incorporating new projects into the pipeline, we provide a dynamic framework that ensures cost constraints remain relevant beyond the initial decision period, addressing the perennial issue of budget allocation in subsequent years.

An innovative aspect of our model is the use of several objective functions. We employed eight different objectives, such as maximizing revenue while adhering to a budget or minimizing costs while achieving a revenue target. This multiplicity allows for a more nuanced understanding of how slight changes in objective functions can significantly impact outcomes. By examining multiple perspectives, decision-makers can combine insights from several optimal solutions to devise a more robust strategy. This approach not only highlights the complex dynamics at play but also underscores that optimal solutions should guide, rather than dictate, strategic decisions.

Our model also introduces constraints designed to shape the portfolio sustainably over time. These constraints include maintaining specific proportions of projects across disease areas and development phases, as well as ensuring a minimum number of new product launches annually. Such constraints help align the portfolio with long-term strategic goals and mitigate risks associated with an imbalanced portfolio. Building on previous work (Wiklund et al., 2023), where different sources for new projects were explored, incorporating these sourcing alternatives into the current model could further enhance its utility by adding another dimension to the decision-making process.

A notable enhancement of our model would be to allow projects to enter the pipeline at any development phase through business development. This adjustment would add realism, as business development and alliances significantly contribute to the pipeline for large pharmaceutical companies (DiMasi et al., 2016). It would also provide a mechanism to quickly address constraints related to project phase proportions, reflecting real-world practices more accurately.

While we have used uniform templates for projects within each category for simplicity, we acknowledge the inherent uniqueness of each drug development project. Future iterations of the model could randomize expected values for timelines and risk-adjusted cash flow curves, better capturing the variability and uncertainty of real projects. This approach, although more complex, would provide a more realistic representation of the portfolio.

An obvious simplification in our current model is the use of expected values for financial metrics, timelines, and future project cash flows. In reality, the net present value (NPV) of drug development projects can vary widely. Projects that fail can result in substantial financial losses, while successful ones can generate significant returns. Incorporating stochastic optimization, where timelines, risks, and cash flows are drawn from distributions rather than fixed expected values, could add a layer of realism and improve decision-making. See e.g., Farid et al. (2021), for an example of stochastic optimization of a pharmaceutical portfolio.

In conclusion, our study presents a novel approach to portfolio optimization by focusing on the inflow of new projects rather than solely on existing ones. This method provides a more dynamic and strategic framework that aligns with the long-term nature of drug development. By using multiple objective functions, and setting future-oriented constraints, our modeling offers a comprehensive tool for decision-makers. Future enhancements, such as incorporating stochastic elements and varying project characteristics, could further refine this approach, making it even more valuable for strategic planning in the pharmaceutical industry.